\documentclass[10pt,conference]{IEEEtran}
\usepackage{mathpple}
\usepackage{mathptmx}
\usepackage{amsmath}
\usepackage{amssymb}
\usepackage{amsbsy}
\usepackage{theorem}
\usepackage{enumerate}
\usepackage{bm}
\usepackage{url}

\newif\ifpdf
\ifx\pdfoutput\undefined
\pdffalse 
\else
\pdfoutput=1 
\pdftrue
\fi

\ifpdf
\usepackage[pdftex]{graphicx}
\usepackage[pdftex,dvipsnames,usenames]{color}
\else
\usepackage{graphicx}
\usepackage[dvips,dvipsnames,usenames]{color}
\fi

\ifpdf
\DeclareGraphicsExtensions{.jpg,.pdf,.tif,.mps,.png}
\else
\DeclareGraphicsExtensions{.eps}
\fi

\graphicspath{{PDF/}}

\usepackage{caption}
\usepackage{subfigure}

\newtheorem{defi}{Definition}
\newtheorem{prop}{Proposition}
\newtheorem{thm}{Theorem}
\newtheorem{lem}{Lemma}
\newtheorem{cor}{Corollary}
\newtheorem{alg}{Algorithm}

\newcommand{\deff}{\ensuremath{\stackrel{\text{\rm def}}{=}}}
\newcommand{\F}{\mathbb{F}}
\newcommand{\proj}[1]{\ensuremath{\!\langle\!\![\,{#1}\,]\!\!\rangle\!}}
\newcommand{\vect}[1]{\ensuremath{\bm{#1}}}

\newcommand{\Eq}[1]{(\ref{#1})}
\newcommand{\Tref}[1]{Theorem \ref{#1}}
\newcommand{\Cref}[1]{Corollary \ref{#1}}
\newcommand{\Lref}[1]{Lemma \ref{#1}}
\newcommand{\AlgRef}[1]{Algorithm \ref{#1}}
\newcommand{\Fref}[1]{Figure \ref{#1}}

\begin{document}

\title{On Algebraic Decoding of $q$-ary Reed-Muller and Product Reed-Solomon Codes}

\author{
\authorblockN{\bf Nandakishore Santhi\authorrefmark{1}}
\authorblockA{
Theoretical Division, CCS-3 Division and the Center for Non Linear Studies\\
LANL, MS B213, T-13, Los Alamos, NM 87545\\
{\tt nsanthi@lanl.gov}
}
}
%

\maketitle

\footnotetext{${}^{\text{\authorrefmark{1}}}$
Document: LA-UR-07-0469.
}

\begin{abstract}
We consider a list decoding algorithm recently proposed by Pellikaan-Wu \cite{PW2005} for $q$-ary Reed-Muller codes $\mathcal{RM}_q(\ell,\; m,\; n)$ of length $n \leq q^m$ when $\ell \leq q$. A simple and easily accessible correctness proof is given which shows that this algorithm achieves a relative error-correction radius of $\tau \leq \left(1 - \sqrt{{\ell q^{m-1}}/{n}}\right)$. This is an improvement over the proof using one-point Algebraic-Geometric codes given in \cite{PW2005}. The described algorithm can be adapted to decode Product-Reed-Solomon codes.

We then propose a new low complexity recursive algebraic decoding algorithm for Reed-Muller and Product-Reed-Solomon codes. Our algorithm achieves a relative error correction radius of $\tau \leq \prod_{i=1}^m \left(1 - \sqrt{k_i/q}\right)$. This technique is then proved to outperform the Pellikaan-Wu method in both complexity and error correction radius over a wide range of code rates.
\end{abstract}

\section{Introduction}
With the discovery of deterministic list-decoding algorithms for several Algebraic-Geometric
codes, most notably the Guruswami-Sudan \cite{GS1999} algorithm, there has been renewed interest in algebraic
decoding methods for other related $q$-ary codes such as the Reed-Muller \cite{PW2004, PW2005}
and Product-Reed-Solomon \cite{PKMV06} codes.
However some of the existing correctness proofs for these algorithms use advanced algebraic geometric
tools. In this paper we first derive a proof for a list decoding algorithm for a $q$-ary Reed-Muller code.
Our proof is from first principles and require only the most basic notions from finite field theory. We then
proceed to propose new recursive list decoding algorithms for Reed-Muller and Product-Reed-Solomon codes. These
algorithms are rigorously shown to outperform the Pellikaan-Wu method in both complexity as well as error-correction-radius.

The basic idea of our new proof for the Pellikaan-Wu algorithm is to ``lift'' a multivariate polynomial in $\F_q[\,x_1,\, x_2,\, \ldots,\, x_m\,]$
to a univariate polynomial in $\F_{q^m}[X]$ using a deterministic mapping rule. This in turn results in a
higher total degree polynomial. The increase in degree will not be high enough to render our list
decoding strategy for Reed-Muller codes useless at meaningful rates. A higher degree for the lifted
polynomial means that this Reed-Muller code list decoding algorithm has a lower relative error-correction radius
(as a function of the rate) than a comparable rate Reed-Solomon list decoder based on the Guruswami-Sudan
algorithm. In the following section we describe the mapping rule and the decoding algorithm in some detail.

In the final section we propose new algorithms for decoding Reed-Muller and Product-Reed-Solomon codes. Our algorithm is more efficient than the Pellikaan-Wu method by approximately a quadratic factor. Furthermore it outperforms the Pellikaan-Wu algorithm in error-correction-radius over a wide range of code rates.

\section{Correctness of a List Decoding Algorithm}
Let us begin by defining a $q$-ary Reed-Muller code.
\begin{defi}
The $q$-ary Reed-Muller code $\mathcal{RM}_q(\ell,\; m,\; n)$ of length $n \leq q^m$ is defined as the set of vectors
given by:
\begin{multline}
\mathcal{RM}_q(\ell,\; m,\; n) \deff \{\;\;[\,\varphi(\bm{\alpha}_1)\; \varphi(\bm{\alpha}_2)\; \cdots\; \varphi(\bm{\alpha}_n)\,]\; \\
  \hspace{2ex}  |\;\;\; \varphi \in \F_q[x_1,\, x_2,\, \ldots,\, x_m],\; \text{\rm deg}(\varphi) \leq \ell\;\;\}
\label{RMEqn}
\end{multline}
where $\{\bm{\alpha}_1, \bm{\alpha}_2, \ldots, \bm{\alpha}_n\}$ are any set of $n$ distinct points in $\F_q^m$.
Here by $\text{\rm deg}(\varphi)$ we mean the total degree of the multivariate polynomial $\varphi$.
\end{defi}

The following well known property will be useful:
\begin{prop}
Let $\{\,a_1,\, a_2,\, \ldots,\, a_m\,\}$
be a basis for $\F_{q^m}$ over $\F_q$ and let $[x_1\, x_2\, \ldots\, x_m] \;\in\; \F_q^m$.
Then the map $\psi : \F_q^m \rightarrow \F_{q^m}$ defined as in \Eq{isomorphEqn} is an isomorphism.
\begin{equation}
[x_1\, x_2\, \ldots\, x_m] \;\mapsto\; X \deff \sum_{j=1}^m a_j x_j
\label{isomorphEqn}
\end{equation}
\end{prop}
For example one might as usual use a polynomial basis
$\{\,1,\, \xi,\, \xi^2,\, \ldots,\, \xi^{m-1}\,\}$ where $\xi$ is any primitive element in $\F_{q^m}$
or even a normal basis of the form $\{\,\zeta,\, \zeta^q,\, \zeta^{q^2},\, \ldots, \zeta^{q^{m-1}}\,\}$,
where $\zeta$ is a suitable primitive element in $\F_{q^m}$.

Therefore we arrive at this elementary conclusion:
\begin{lem}
Let $X \in \F_{q^m}$. The reverse isomorphism for \Eq{isomorphEqn} is:
\begin{equation}
X \;\mapsto\; [x_1\, x_2\, \ldots\, x_m]^T \deff \vect{A}^{-1}\cdot [X\; X^q\; X^{q^2} \;\ldots\; X^{q^{m-1}}]^T
\label{revIsoEqn}
\end{equation}
where
\begin{equation}
\vect{A}
\deff
\begin{bmatrix}
a_1 & a_2 & \ldots & a_m \\
a_1^q & a_2^q & \ldots & a_m^q \\
\vdots & \vdots & \ddots & \vdots \\
a_1^{q^{m-1}} & a_2^{q^{m-1}} & \ldots & a_m^{q^{m-1}}
\end{bmatrix}
\label{AmatrixEqn}
\end{equation}
is a non-singular (invertible) square matrix.
\label{revIso}
\end{lem}
\proof
Since $X = \sum_{j=1}^m a_j x_j$, and $x_j \in \F_q$, we get $X^{q^i} = \sum_{j=1}^m a_j^{q^i} x_j$
using Fermat's little theorem. It only remains to show that $\vect{A}$ is non-singular. Note that in general $\vect{A}$ is \emph{not} a Vandermonde matrix. However by construction, the set
$\{\,a_1,\, a_2,\, \ldots,\, a_m\,\}$ is a basis for $\F_{q^m}$ over $\F_q$. It then follows from
\cite[Corollary 2.38, pp. 58]{NL1986} that $\vect{A}$ is non-singular.
\QED
\vspace*{1ex}

It follows from \Lref{revIso} that there exist polynomials $\mu_j\in\F_{q^m}[X]$ of degree at most $q^{m-1}$ such that
$x_j = \mu_j(X), 1 \leq j \leq m$. Substituting for all $x_j$ in this manner, we have proved the following:
\begin{thm}
Let $n \leq q^m$. If $\ell \leq q$ then
\begin{equation}
\mathcal{RM}_q(\ell,\; m,\; n) \subseteq \mathcal{RS}_{q^m}(n,\; \ell q^{m-1}) \cap \F_q^n
\label{RScontRMEqn}
\end{equation}
where $\mathcal{RS}_{q^m}(n,\; \ell q^{m-1})$ is the Reed-Solomon code given by
\begin{multline}
\mathcal{RS}_{q^m}(n,\; \ell q^{m-1}) \deff \{\;\;[\,f(\beta_1)\; f(\beta_2)\; \ldots\; f(\beta_n)\,] \; \\
\hspace{2ex} |\;\;\; f\in\F_{q^m}[X], \;\text{\rm deg}(f) \leq \ell q^{m-1}\;\;\}
\end{multline}
where $\beta_i \deff \sum_{j=1}^m a_j \alpha_{ij}$, and $\bm{\alpha}_i \deff [\alpha_{i1} \;\alpha_{i2}\; \ldots\; \alpha_{im}]$, $1\leq i\leq n$ are the points of evaluation for the Reed-Muller code. Moreover if the information polynomial associated with the Reed-Muller code is given by
\begin{equation}
\varphi(\,x_1,\, x_2,\, \ldots,\, x_m\,) \deff \sum_{\stackrel{i_1,i_2,\ldots,i_m \, :}{\sum_j i_j \leq \ell}}\varphi_{i_1,i_2,\ldots,i_m}\prod_{j=1}^m x_j^{i_j}
\label{RMinfoPolyEqn}
\end{equation}
then the information polynomial $f$ of degree at most $\ell q^{m-1}$ associated with the Reed-Solomon code is:
\begin{equation}
f(X) = \sum_{\stackrel{i_1,i_2,\ldots,i_m \, :}{\sum_j i_j \leq \ell}}\varphi_{i_1,i_2,\ldots,i_m}\prod_{j=1}^m\left(\mu_j(X)\right)^{i_j}
\label{RSinfoPolyEqn}
\end{equation}
\label{embedThm}
\end{thm}

Let $d_H(\vect{x},\; \vect{y})$ represent the Hamming distance between
the two vectors.
Using \Tref{embedThm} and the Guruswami-Sudan algorithm \cite{GS1999} for list decoding a Reed-Solomon code,
we have proved the correctness of the following deterministic list-decoding algorithm for Reed-Muller codes:
\begin{alg}[RM-List-1]
\hspace{0ex}\\
\underline{\sc{\bf{INPUT}}:} $q, \ell\leq q, m, n\leq q^m$; $\vect{r} = [\,r_1\; r_2\; \ldots\; r_n\,] \in \F_q^n$.\\
\underline{\sc{\bf{STEPS}}:}
\begin{enumerate}[{\bf 1.}]
\item
Compute the parameter $t = \left\lceil n\left(1 - \sqrt{{\ell q^{m-1}}/{n}}\right)\right\rceil$.
\item
Using Guruswami-Sudan algorithm find a list $\mathcal{L}$ of codewords $\vect{c} \in \mathcal{RS}_{q^m}(n,\; \ell q^{m-1})$ such that
$
d_H(\vect{c},\; \vect{r}) < t
$.
\item
For every $\vect{c} \in \mathcal{L}$ check if $\vect{c} \in \F_q^n$ {\rm :}
\begin{enumerate}[{\bf i.}]
\item
If {\sc no} then discard $\vect{c}$ from $\mathcal{L}$.
\item
If {\sc yes} then check if $\vect{c} \in \mathcal{RM}_q(\ell,\; m,\; n)$ {\rm :}
\begin{enumerate}[{\bf a.}]
\item
If {\sc no} then discard $\vect{c}$ from $\mathcal{L}$.
\item
If {\sc yes} then keep $\vect{c}$ in the list $\mathcal{L}$.
\end{enumerate}
\end{enumerate}
\item \underline{\tt return}
\end{enumerate}
\underline{\sc{\bf{OUTPUT}}:} $\mathcal{L}$
\label{mainAlgo}
\end{alg}
This algorithm was originally proposed by Pellikaan-Wu in \cite{PW2005}, though their proofs were different.

\subsection{Complexity of \AlgRef{mainAlgo}}
The complexity of our proposed algorithm is of the same order as the complexity
of Guruswami-Sudan algorithm for decoding Reed-Solomon codes over the extension
field $\F_{q^m}$. This is $\mathcal{O}(n^3)$ field operations in $\F_{q^m}$.

\subsection{Comparison to previous results}
The Pellikaan-Wu algorithm for decoding Reed-Muller codes by means of embedding into one-point Algebraic-Geometric codes
was shown \cite{PW2005} to achieve an error correction radius of
$\left\lceil n\left(1 - \sqrt{{\ell (q+1)^{m-1}}/{n}}\right)\right\rceil$.
It is interesting to note that the error-correction radius demonstrated herein is always larger than that suggested by the Pellikaan-Wu
formalism employing Algebraic-Geometric codes. However we believe that the more important contribution
of this paper is the readily accessible correctness proof which relies on just a few basic notions from Galois theory.

\subsection{Product Reed-Solomon codes}
Product Reed-Solomon codes $\mathcal{PRS}_{q,m}(q^m,\; k_1,\,\ldots,\,k_m) \deff \otimes_{i=1}^m\mathcal{RS}_q(q,\; k_i)$
over $\F_q^m$ can be thought of as the set of vectors whose $q^m$ coordinates consist of the $q^m$
evaluations of $m$-variate
information polynomials with coefficients in $\F_q$ and degree in the $i^{th}$-variable $x_i$ at most $(k_i-1)$. $m$ is usually called the dimension of the product code. Thus
$\mathcal{PRS}_{q,m}(q^m,\; k_1,\,\ldots,\,k_m)$ is contained in $\mathcal{RM}_q(\sum_{i=1}^m (k_i-1),\; m,\; q^m)$.
When $\sum_{i=1}^m (k_i-1) \leq q$ the list decoding
algorithm given in \AlgRef{mainAlgo} may be used essentially without any modifications. Several Product-Reed-Solomon
algebraic list decoders, including a similar method as sketched above are described in \cite{PKMV06}. Using \AlgRef{mainAlgo}
it is possible to achieve a relative error correction radius of $(1 - \sqrt{\sum_{i=1}^m \rho_i})$, where $\rho_i \deff k_i/q$.

\subsection{Zeros of Multivariate Polynomials}
From \Tref{embedThm}, it is clear that $f(X)$ being of degree at most $\ell q^{m-1}$, has at most $\ell q^{m-1}$ zeros in $\F_{q^m}$, including
multiplicities.
Therefore a non-zero multivariate polynomial $\varphi(\,x_1,\, x_2,\, \ldots,\, x_m\,)$ of total degree $\ell$ has at most $\ell q^{m-1}$
zeros in $\F_q^m$. This gives the famous DeMillo-Lipton-Schwartz-Zippel\cite{Sch80} lemma for polynomials over finite fields. Note that the statement above
appears to be stronger than the classical lemma in that this counts multiplicities too. Moreover the proof also appears to differ from the traditional
expositions which use probabilistic arguments.

Next we propose a lower complexity recursive algebraic decoder which outperforms the Reed-Muller decoder considered in this section.

\section{A Recursive Decoding Algorithm for Reed-Muller and Product Reed-Solomon Codes}

\begin{figure}[!htbp]
\vspace*{3ex}
\centerline{\includegraphics[width=0.48\textwidth,keepaspectratio]{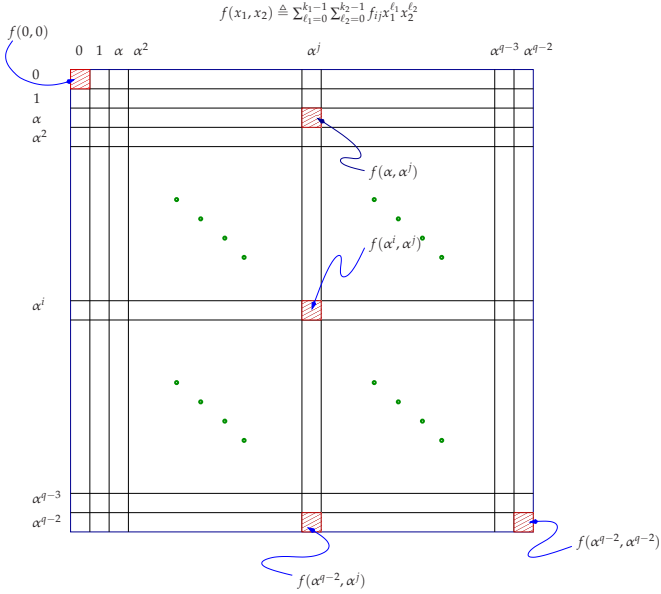}}
\caption{2D PRS code described on a rectangular array.\label{2DPRS}}
\end{figure}

For simplicity, let $n \deff q^m$. A codeword in the code $\mathcal{PRS}_{q,m}(q^m,\; k_1,\,\ldots,\,k_m)$ can be described within an $m$-dimensional cube of side length $q$. See \Fref{2DPRS}. Let a codeword $\vect{c}$ (correspondingly a received word, $\vect{r}$) be so described. We will find it convenient to write this vector as $\proj{\vect{c}_{i_1, i_2, \ldots, i_m}}$, where each of the indices $i_j$ take values in the range $\{1, \ldots, q\}$. We further use the notation $\proj{\vect{c}_{i_1, i_2, \ldots, i_{j-1}}^{a_j, a_{j+1}, \ldots, a_m}}$ to denote the $(j-1)$-dimensional vector formed out of $\proj{\vect{c}_{i_1, i_2, \ldots, i_m}}$ when the coordinates indexed by $(i_j, i_{j+1}, \ldots, i_m)$ are fixed at $(a_j, a_{j+1}, \ldots, a_m)$ and the rest of the indices are free. By the nature of the product code, $\proj{\vect{c}_{i_1, i_2, \ldots, i_{j-1}}^{a_j, a_{j+1}, \ldots, a_m}}$ belongs to $\mathcal{PRS}_{q,j-1}(q^{j-1},\; k_1,\,\ldots,\,k_{j-1})$.

Now consider the following decoding algorithm for the code $\mathcal{PRS}_{q,m}(q^m,\; k_1,\,\ldots,\,k_m)$:
\begin{alg}[PRS-Decoder]
\hspace{0ex}\\
\underline{\sc{\bf{INPUT}}:} $q, (k_1, k_2, \ldots, k_m): k_i<q, m$; $\vect{r} \in \F_q^n$, where $\vect{r} \deff \langle\![\,r_{i_1, i_2, \ldots, i_m}\,]\!\rangle; 1 \leq i_j \leq q$.\\
\underline{\sc{\bf{STEPS}}:}
\begin{enumerate}[{\bf 1.}]
\item
If $m=1$ do:
\begin{enumerate}[{\bf i.}]
\item
Compute the parameter $t_1 = \left\lceil q\left(1 - \sqrt{k_1/q}\right)\right\rceil$.
\item
Using Guruswami-Sudan algorithm find a list $\mathcal{L}_1$ of codewords $\vect{c}_1 \in \mathcal{RS}_q(q,\; k_1)$ such that
$
d_H(\vect{c}_1,\; \proj{\vect{r}_{i_1}}) < t_1
$.
\item
Search $\mathcal{L}_1$ for $\vect{c}_1$ such that $d_H(\vect{c}_1,\; \proj{\vect{r_{i_1}}})$ is least. Substitute in-place
the positions corresponding to $\proj{\vect{r_{i_1}}}$ in $\vect{r}$ with $\vect{c}_1$ and \underline{\tt return}.
\end{enumerate}
\item
For $a_m = 1, 2, \ldots, q$ do:
\begin{enumerate}[{\bf i.}]
\item
Set $\vect{r}' \leftarrow
\proj{\vect{r}_{i_1, i_2, \ldots, i_{m-1}}^{a_m}}$
\item
Set $m' \leftarrow m-1$ and $n' \leftarrow q^{m'}$
\item
Recursively decode $\vect{r}'$ using {\bf PRS-Decoder} with input parameters $q, (k_1, k_2, \ldots, k_{m'}), m'$; $\vect{r}' \in \F_q^{n'}$. 
\end{enumerate}

\item
Compute the parameter $t_m = \left\lceil q\left(1 - \sqrt{k_m/q}\right)\right\rceil$.
\item
For each $m-1$ tuple $(a_1, a_2, \ldots, a_{m-1})$ do:
\begin{enumerate}[{\bf i.}]
\item
Using Guruswami-Sudan algorithm find a list $\mathcal{L}_m$ of codewords $\vect{c}_m \in \mathcal{RS}_q(q,\; k_m)$ such that
$
d_H(\vect{c}_m,\; \proj{\vect{r_{i_m}^{a_1, a_2, \ldots, a_{m-1}}}} < t_m
$.
\item
Search $\mathcal{L}_m$ for $\vect{c}_m$ such that $d_H(\vect{c}_m,\; \proj{\vect{r_{i_m}^{a_1, a_2, \ldots, a_{m-1}}}})$ is least. Substitute in-place
the positions corresponding to $\proj{\vect{r_{i_m}^{a_1, a_2, \ldots, a_{m-1}}}}$ with $\vect{c}_m$.
\end{enumerate}
\item \underline{\tt return}
\end{enumerate}
\underline{\sc{\bf{OUTPUT}}:} Resulting vector $\vect{r}$
\label{recPRSAlgo}
\end{alg}

The following recursive algorithm uses {\bf PRS-Decoder} to decode $\mathcal{RM}_q(\ell,\; m,\; n)$.
\begin{alg}[RM-List-2]
\hspace{0ex}\\
\underline{\sc{\bf{INPUT}}:} $q, \ell\leq q, m, n\leq q^m$; $\vect{r} = [\,r_1\; r_2\; \ldots\; r_n\,] \in \F_q^n$.\\
\underline{\sc{\bf{STEPS}}:}
\begin{enumerate}[{\bf 1.}]
\item
For each possible $m$-tuple $(k_1, k_2, \ldots, k_{m}): k_i<q, \sum_j k_j \leq \ell$ do:
\begin{enumerate}[{\bf i.}]
\item
Using {\bf PRS-Decoder} with input parameters
$q, (k_1, k_2, \ldots, k_m), m$; $\vect{r} \in \F_q^n$, decode $\vect{r}$ as $\vect{c}$.
\item
Add $\vect{c}$ to a list $\mathcal{L}$ of codeword candidates.
\end{enumerate}
\item \underline{\tt return}
\end{enumerate}
\underline{\sc{\bf{OUTPUT}}:} $\mathcal{L}$
\label{recRMAlgo}
\end{alg}

We have the following result concerning the decoding power of \AlgRef{recPRSAlgo} and \AlgRef{recRMAlgo}.
\begin{thm}
\AlgRef{recPRSAlgo} has a relative error correction radius of $\tau_m \deff \prod_{i=1}^m (1-\sqrt{\rho_i})$, where $\rho_i \deff k_i/q$. Moreover, there exist error patterns of weight above $n \prod_{i=1}^m (1-\sqrt{\rho_i})$ which cannot be guaranteed to be efficiently decoded by \AlgRef{recPRSAlgo}.
\label{recAlgoThm}
\end{thm}
\proof
\begin{figure}[!htbp]
\vspace*{3ex}
\centerline{\includegraphics[width=0.48\textwidth,keepaspectratio]{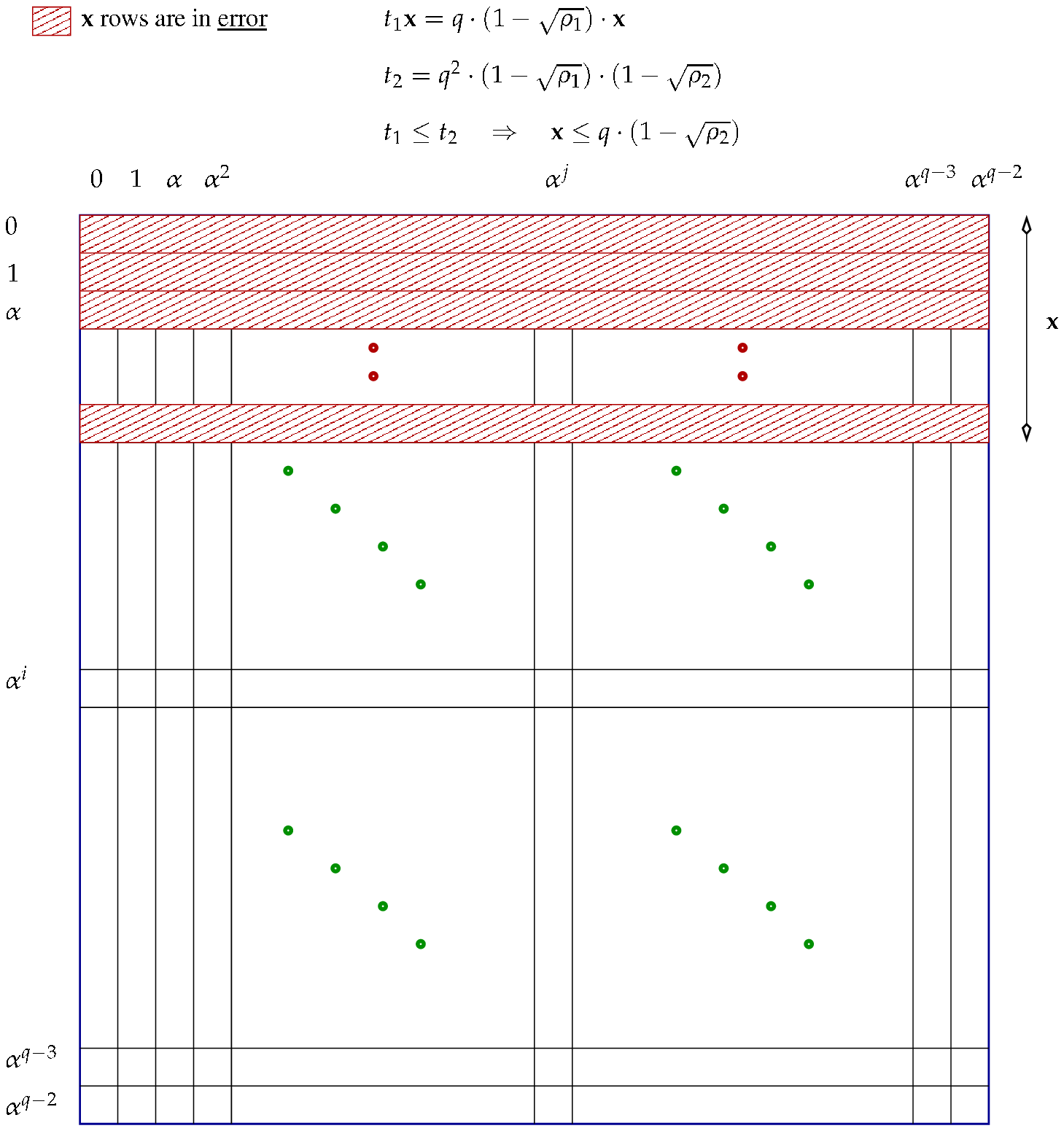}}
\caption{The proof of \Tref{recAlgoThm} for a 2D PRS code.\label{2DPRSrad}}
\end{figure}

Our proof is by induction. When $m=1$, the claim is trivially true. Let us assume the claim to be true for some $m=M$. We will now show it to be true for the case $m=M+1$. Let there be a maximum of $t_{M+1} = q^{M+1} \prod_{i=1}^{M+1} (1-\sqrt{\rho_i})$ errors. In Step 2 of \AlgRef{recPRSAlgo}, 
let there be a maximum of $x$ recursions which fail to decode correctly. Since by the induction hypothesis, this would mean that there are more than $t_M$ errors in these $x$ sub-recursions, we have that $x\, t_M \leq t_{M+1}$. Substituting for $t_{M+1}$ and $t_M$ gives, $x \leq q(1-\sqrt{\rho_{M+1}})$. These errors will get corrected in Step 4 of the algorithm. This proves the first part of the claim. For the 2D case, the proof is concisely depicted in \Fref{2DPRSrad}.

To see the second part of the claim, we observe that an error pattern which is contiguously spread over an $m$ dimensional sub-cube of volume more than $n \prod_{i=1}^m (1-\sqrt{\rho_i})$ cannot be guaranteed to be efficiently decoded by the proposed algorithm. This shows that the error correction radius predicted in the first part of \Tref{recAlgoThm} is rather tight.
\QED

\subsection{Complexity of \AlgRef{recPRSAlgo} and \AlgRef{recRMAlgo}}
Let $\vartheta_m$ be the complexity of decoding an $m$-dimensional Product-Reed-Solomon code using \AlgRef{recPRSAlgo}. Then the complexity of decoding an $m+1$ dimensional code is $\vartheta_{m+1} = \mathcal{O}(q\, \vartheta_m \,+\, q^m\, \vartheta_1)$. But $\vartheta_1 = \mathcal{O}(q^3)$ field operations in $\F_q$. This gives, $\vartheta_m = \mathcal{O}(q^{m+2})$ which is $\approx \mathcal{O}(n)$ for large $m$. The complexity of \AlgRef{recRMAlgo} is $\approx \mathcal{O}(n^2)$ field operations in $\F_q$. This is substantially better than the Pellikaan-Wu method in \AlgRef{mainAlgo}.

\subsection{Comparison of \AlgRef{mainAlgo} and \AlgRef{recRMAlgo}}
\begin{figure}[!htbp]
\centering
\subfigure[Relative error correction radii of the two algorithms are compared when decoding a 2D Product-Reed-Solomon code. The surface which dominates for most of the rate region corresponds to the new algorithm.\label{RMperf}]{\includegraphics[width=0.4\textwidth,keepaspectratio]{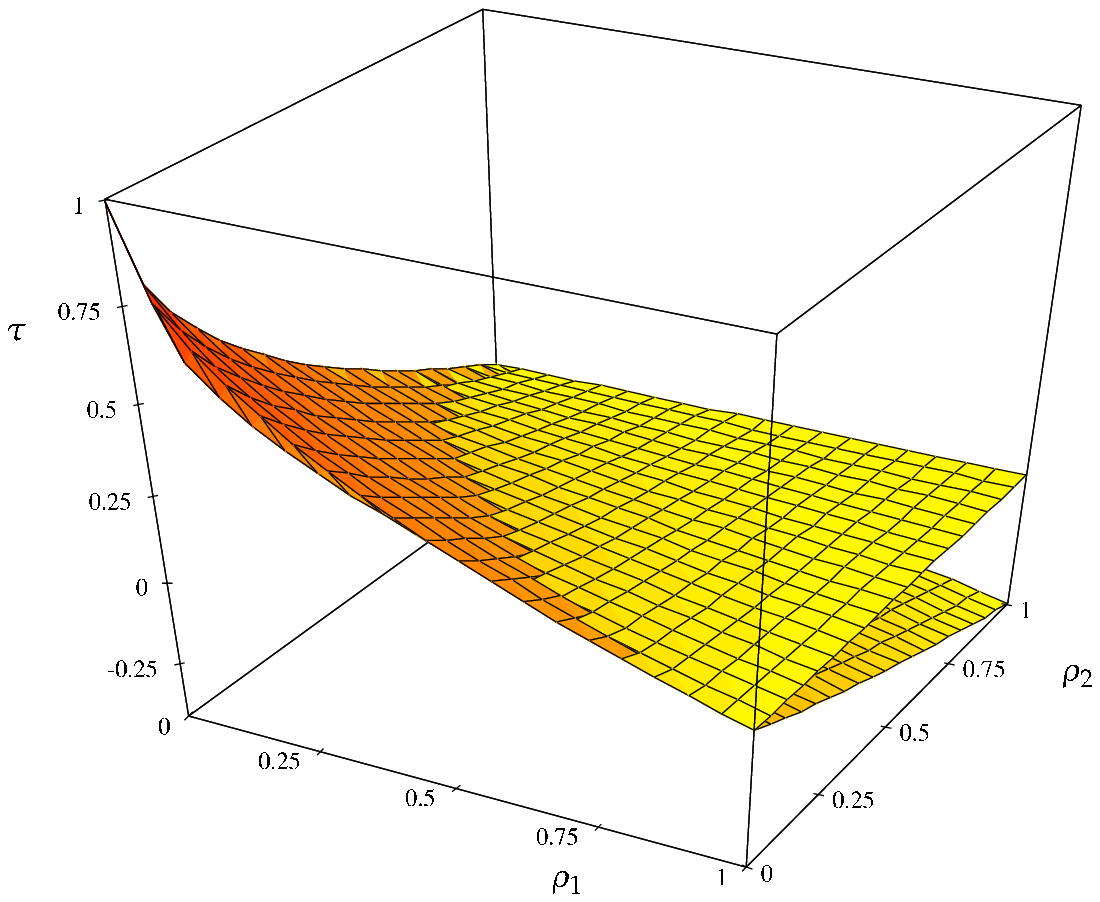}}
\subfigure[2D rate region where the new recursive algorithm performs better is shown.\label{V2pie}]{\includegraphics[width=0.28\textwidth,keepaspectratio]{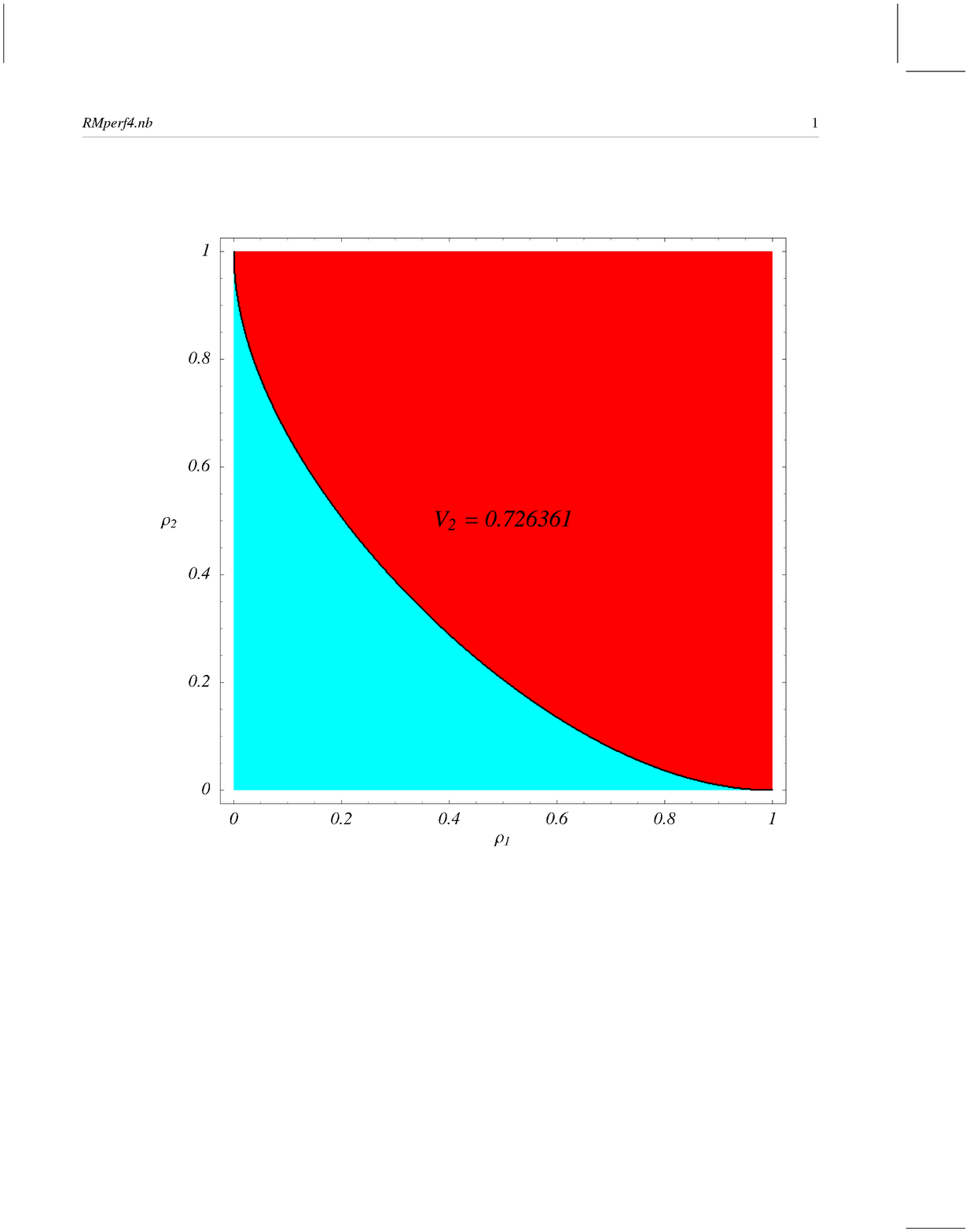}}
\subfigure[Relative rate region volume where our algorithm performs better is computed for various Product-Reed-Solomon codes of dimensionality $m$. The new algorithm is seen to out-perform the Pellikaan-Wu algorithm over much of the rate region.\label{VmPlot}]{\includegraphics[width=0.4\textwidth,keepaspectratio]{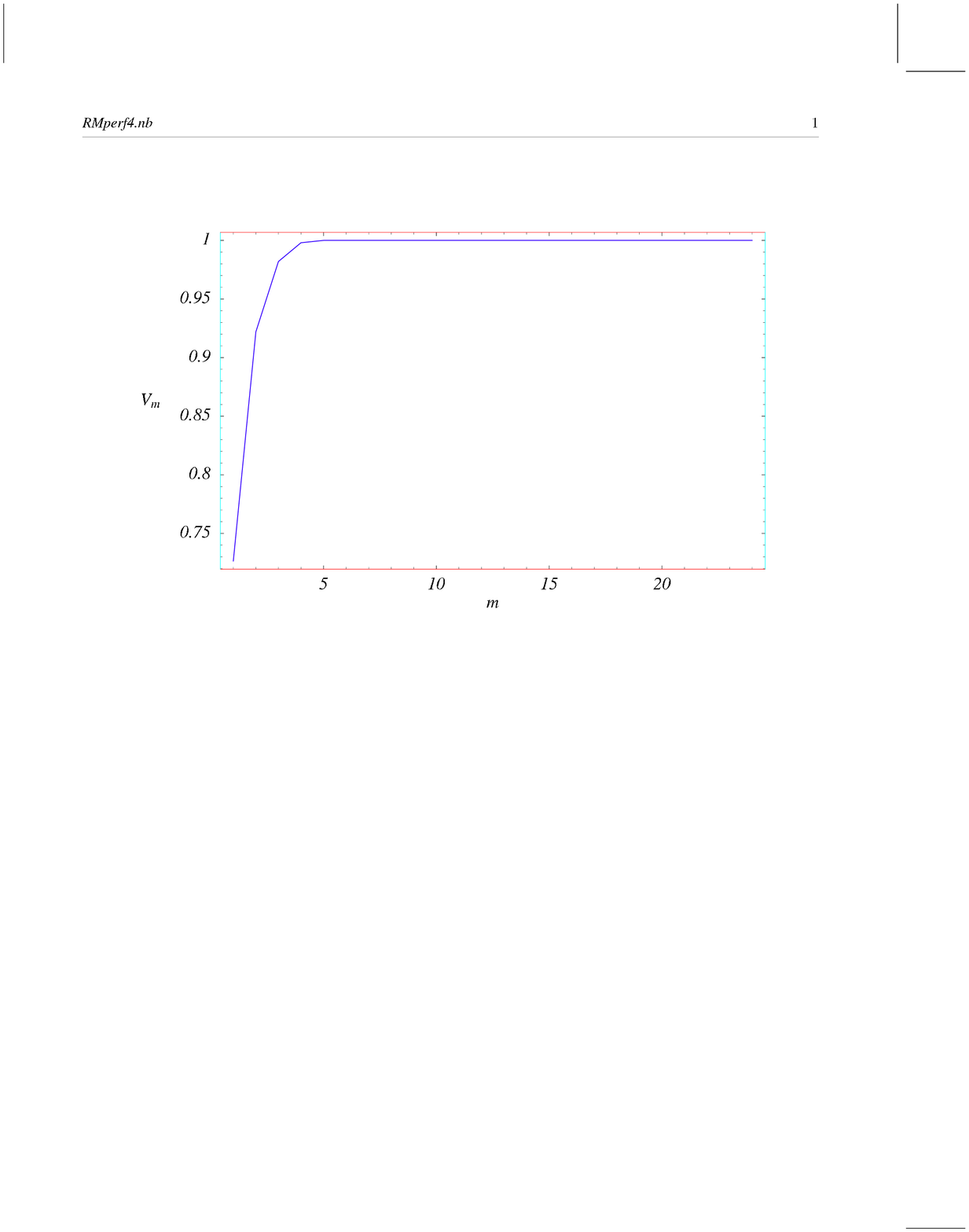}}
\caption{Comparison of Pellikaan-Wu algorithm to our new recursive RM/PRS decoder.\label{PWSanthiCompare}}
\end{figure}

\AlgRef{recRMAlgo} not only has a lower complexity, but also performs better over a wide range of rates. For example when $\sum_i \rho_i > 1$, the Pellikaan-Wu algorithm is not effective, whereas the new algorithm is still useful. Furthermore $\prod_{i=1}^m (1-\sqrt{\rho_i})$ is larger than $(1-\sqrt{\sum_{i=1}^m {\rho_i}})$ for most code rates and the advantage is more pronounced at higher code rates. \Fref{PWSanthiCompare} shows the decoding power of \AlgRef{recPRSAlgo}.

\subsection{Other Related Product Code Decoders}
Several iterative hard decision decoders for Product-Reed-Solomon codes available in literature use some form of \AlgRef{recPRSAlgo}. Usually such algorithms are described with no theoretical bounds on their error correction radii. These product code decoders find use in optical communication systems and LAN/WAN standards\cite{IEEE802, IEEE802.16}. Several hardware implementations of such decoders are commercially available\cite{AHA,XILINX,ALTERA}. Soft decision iterative decoders for product codes utilizing the ''turbo-principle'' have also been discussed in literature\cite{Pyn98}. The performance of most of these hard decision iterative decoders can be very well characterized using \Tref{recAlgoThm}. Similar conclusions are obvious for the case of other product codes which have algebraic bounded distance decoders available for their component codes. \Tref{recAlgoThm} implies the following for a general product code:
\begin{cor}
If for an $m$-dimensional product code $\mathbb{P}$, there exists bounded distance decoders for each of its component codes such that the $i^{th}$ component code's decoder achieves a error correction radius of $t_i$ errors, then there exists a decoding algorithm for the entire product code $\mathbb{P}$ which can correct all errors up to a weight of $t = \prod_{i=1}^m{t_i}$.
\label{recAlgoCor}
\end{cor}
The decoding algorithm for the code $\mathbb{P}$ mentioned in \Cref{recAlgoCor} can be obtained from \AlgRef{recPRSAlgo} with some obvious and minor changes and as such is not repeated here. This result is, to the best of the author's knowledge, the only such theoretical guarantee on the error correction radius of a general algebraic product code decoder. However for specific cases there are some stronger results available, for instance see the result of Lin-Weldon\cite{LW70} for cyclic product codes. In another related example, Tanner\cite{Tan81} discusses bounds on a specific type of hard-decision decoder for product codes on graphs. In many cases of practical interest such bounds are difficult to apply because of their dependence on the knowledge of the girth of the underlying code graph.

\section{Conclusions}
In this paper, we presented a simple and easily accessible proof for the Pellikaan-Wu algebraic list decoding algorithm for Reed-Muller codes. Our proof uses only the most fundamental properties of finite field arithmetic.
\newpage
We also proposed a low complexity recursive algorithm for Reed-Muller and Product-Reed-Solomon codes. This new recursive algebraic decoding algorithm is then shown to have a significantly better error correction radius than the Pellikaan-Wu algorithm over a wide range of code rates.

%



\end{document}